\documentclass[12pt,a4paper,dvips]{article}
\usepackage{a4p}
\usepackage{cite,mcite}
\usepackage{graphicx}
\usepackage{rotating,a4p,here}
\usepackage{phy_susy, amssymb}
\usepackage{amsmath}
\usepackage{l3_titleppe,ifthen}
\journalname{To be submitted to Phys. Lett. B}
\preprint{01-068}
\date{October 9, 2001}
\newlength{\capindent}
\newlength{\capwidth}
\newlength{\figwidth}
\newcommand{\icaption}[2][!*!,!]{\hspace*{\capindent}%
  \begin{minipage}{\capwidth}
    \ifthenelse{\equal{#1}{!*!,!}}%
      {\caption{#2}}%
      {\caption[#1]{#2}}
  \end{minipage}}
\newcommand{\pho}{\phantom{0}}
\newcommand{\phopunto}{\phantom{.0}}
\newcommand{\phouno}{\phantom{1}}

\newcommand{\rpvpha}{\vphantom{$\overline{\slep_j}$}}

\def\lamtre{\ifmath{\lambda_{133}}}
\def\lamuno{\ifmath{\lambda''_{212}}}

\def\ycut{\ifmath{y_{cut}}}

\def\Emiss{\ensuremath{E\hspace{-.23cm}/\hspace{+.01cm}}}
\def\LLE{\ifmath{\mathrm{L}_{i}\mathrm{L}_{j}{\overline{\mathrm E}_{k}}}}
\def\LQD{\ifmath{\mathrm{L}_{i}\mathrm{Q}_{j}{\overline{\mathrm D}_{k}}}}
\def\UDD{\ifmath{{\overline{\mathrm U}_{i}}{\overline{\mathrm D}_{j}}{\overline{\mathrm D}_{k}}}}
\def\sq{\ifmath{\tilde{\rm  q}}}

\def\snu{\ifmath{\tilde{\nu}}}
\def\snue{\ifmath{\tilde{\nu}_e}}

\def\snumt{\ifmath{\tilde{\nu}_{\mu,\tau}}}
\def\slep{\ifmath{\tilde{\ell}}}
\def\slepp{\ifmath{\tilde{\ell}^+_R}}
\def\slepm{\ifmath{\tilde{\ell}^-_R}}
\def\slepr{\ifmath{\tilde{\ell}_R}}
\def\serr{\ifmath{\tilde{\rm e}_R}}
\def\serp{\ifmath{\tilde{\rm e}^+_R}}
\def\serm{\ifmath{\tilde{\rm e}^-_R}}
\def\smur{\ifmath{\tilde{\mu}_R}}
\def\smurp{\ifmath{\tilde{\mu}^+_R}}
\def\smurm{\ifmath{\tilde{\mu}^-_R}}
\def\staur{\ifmath{\tilde{\tau}_R}}
\def\staurp{\ifmath{\tilde{\tau}^+_R}}
\def\staurm{\ifmath{\tilde{\tau}^-_R}}
\def\qur{\ifmath{\tilde{\rm u}_R}}
\def\qul{\ifmath{\tilde{\rm u}_L}}
\def\qdr{\ifmath{\tilde{\rm d}_R}}
\def\qdl{\ifmath{\tilde{\rm d}_L}}
\def\susy#1{\ensuremath{\tilde{\mathrm{#1}}}}%

\def\neutralino#1{\ensuremath{\susy{\chi}_{#1}^0}}

\def\mo{\ensuremath{m_0~}}

\def\tb{\ensuremath{\tan\beta}}

\begin{document}
\begin{titlepage}
\title{ \boldmath{Search for R-parity Violating Decays of Supersymmetric
       Particles in $\mathrm{e^+e^-}$ Collisions at LEP}}  

\author{The L3 Collaboration}
%
%
\begin{abstract}
A search,  in $\mathrm \epem$ collisions,
for chargino, neutralino, scalar lepton and scalar quark
pair-production is performed, without assuming R-parity
conservation in decays, in the case that only one of the 
coupling constants $\lambda_{ ijk}$ or $\lambda''_{ ijk}$
is non-negligible.
No signal is found in data up to a centre-of-mass energy
of 208 \GeV. 
Limits on the production cross sections
and on the masses of supersymmetric particles  
are derived.

\end{abstract}

\submitted
\end{titlepage}

\section{Introduction}

The most general superpotential of the
Minimal Supersymmetric Standard Model (MSSM) ~\cite{MSSM}, which describes
a supersymmetric, renormalizable
and gauge invariant theory, with minimal particle content, includes
the term $\mathrm{W_R}$~\cite{superp,superp2}:
\begin{equation} \mathrm{W_R} =
    \lambda_{ ijk} \mathrm L_i \mathrm L_j \overline{\mathrm E}_k \, +
    \lambda'_{ ijk} \mathrm L_i \mathrm Q_j \overline{\mathrm D}_k \, +
    \lambda''_{ ijk} \overline{\mathrm U}_i \overline{\mathrm D}_j 
                            \overline{\mathrm D}_k\,,
\label{eqn:wr}
\end{equation}
where $\lambda_{ ijk}$, $\lambda'_{ ijk}$ and $\lambda''_{ ijk}$ 
denote the Yukawa couplings
and {\it i, j} and {\it k} the generation indices;
$\mathrm L_i$ and $\mathrm Q_i$ are the left-handed lepton- and 
quark-doublet superfields,
$\overline{\mathrm E}_i$, $\overline{\mathrm D}_i$ and $\overline{\mathrm U}_i$
are the right-handed singlet superfields for
charged leptons, down- and up-type quarks, respectively.
The $\LLE$ and $\LQD$ terms violate the leptonic quantum number L, while 
the $\UDD$ terms violate the baryonic quantum number B.

R-parity is a multiplicative quantum number defined as:
\begin{equation} \mathrm
    R = (-1)^{\mathrm {3B+L+2S}}\,,
\end{equation}
where~S~is~the~spin. For ordinary particles~R~ is $+1$,~while it 
is~$-1$~for~their~supersymmetric partners.
R-parity conservation implies that supersymmetric particles can only be
produced in pairs and then decay in cascade to the lightest supersymmetric
particle (LSP), which is stable~\cite{rpc_susy1}.
This hypothesis is formulated in order to prevent a fast proton 
decay~\cite{weinberg}, disfavoured by present limits~\cite{pdg2000}.
However, the absence of either the B- or the L-violating terms
is enough to prevent such a decay, and the hypothesis of 
R-parity conservation can be relaxed.
As a consequence, two new kinds of processes are allowed:
single production of supersymmetric 
particles~\cite{single_snu, rpv_publ1}, or LSP
decays into Standard Model particles
via scalar lepton or scalar quark exchange. For these decays, the MSSM
production mechanisms are unaltered by the operators in 
Equation~\ref{eqn:wr}. In this letter, the cases in which either a
neutralino or a scalar lepton is the LSP are considered.

In this paper, we describe the
search for pair-produced neutralinos
($\mathrm \epem \ra \chim\chin$,
with $m = 1,2 \,$ and $n = 1,..,4$), 
charginos ($\mathrm \epem \ra \chap\cham$),
scalar leptons ($\mathrm \epem \ra \slepp \slepm$, where
$\tilde{\ell}^\pm_R$ represents scalar electrons, muons or tau and
$\epem \ra \snu\snu$)
and scalar quarks ($\mathrm \epem \ra \sq\sq$)
with subsequent R-parity violating decays, assuming that only
one of the coupling constants $\lambda_{ ijk}$
or $\lambda''_{ ijk}$ is non-negligible.
Only the supersymmetric partners of the right-handed charged leptons,
$\slepr$, are considered, as they are expected to be lighter
than the corresponding left-handed ones.

Supersymmetric particles can either decay directly into two or 
three fermions according to the dominant interaction term,  
or indirectly via the LSP. The different decay modes are detailed in 
Table~\ref{tab:decays}.
Four-body decays of the lightest scalar lepton are also taken into
account in the case of $\lambda''_{ ijk}$.
In the present analysis, the dominant coupling is assumed to be greater than
$10^{-5}$~\cite{dawson}, which corresponds to decay lengths below 1 cm.

Previous L3 results at centre-of-mass energies ($\sqrt s$) 
up to 189 \GeV{}
are reported in References~\citen{rpv1_l3} and~\citen{rpv2_l3},
where also $\lambda'_{ijk}$ couplings are discussed. 
Two new analyses are presented in this letter: 
$\epem \ra \snu\snu$ and $\mathrm \epem \ra\sq\sq$ in the case of
$\lambda''_{ijk}$ couplings. New interpretations for scalar leptons and
scalar quarks in the MSSM framework are also performed.

Searches for R-parity violating decays of supersymmetric particles
were also reported by other LEP experiments~\cite{rpv_publ1, rpv_publ2}.

\begin{table*} [htbp] 
  \begin{center}
  \begin{tabular}{|l|c|c|c|c|} \hline 

   \multicolumn{1}{|c|} {Particle}
   &\multicolumn{2}{c|} {Direct decays }
   &\multicolumn{2}{|c|} {Indirect decays } \\ \cline{2-5}

   \multicolumn{1}{|c|} { }
   &\multicolumn{1}{c|} {$\lambda_{ ijk}$}
   &\multicolumn{1}{c|} {$\lambda''_{ ijk}$}
   &\multicolumn{1}{|c|} { via $\neutralino{1}$}
   &\multicolumn{1}{|c|} { via \vphantom{$\overline{\slep_j}$} 
                          $\slep$} \\ \hline

     $\chio$  \vphantom{$\overline{\slep_j}$}
     & $ \ell_i^- \nu_j \ell_k^+$, $\nu_i \ell^+_j \ell^-_k$  
     & $ \bar{\mathrm u}_i \bar{\mathrm d}_j \bar{\mathrm d}_k$  
     &  $-$   
     & $\ell\slep$ \\ \hline

    $\neutralino{n (n \ge 2)}$   \vphantom{$\overline{\slep_j}$}
     & $ \ell_i^- \nu_j \ell_k^+$, $\nu_i \ell^+_j \ell^-_k$  
     & $ \bar{\mathrm u}_i \bar{\mathrm d}_j \bar{\mathrm d}_k$  
       &  $\Zstar\chim_{(m < n)}$, 
     & $\ell\slep$ \\ 

      {  }
     & 
     & 
     & $\Wstar\cha$  
     & \\ \hline

    $\chap$   \vphantom{$\overline{\slep}_j$}
     & $ \nu_i \nu_j \ell^+_k$, $\ell^+_i \ell^+_j \ell^-_k$
     & $\bar{\mathrm d}_i \bar{\mathrm d}_j \bar{\mathrm d}_k$, 
       $\mathrm u_i \mathrm u_j \mathrm d_k$,
     & $\Wstar\chio$, $\Wstar\chid$
     &  \\
 
        { }
     & 
     & $\mathrm u_i \mathrm d_j \mathrm u_k$ 
     & 
     &  \\ \hline

     $\slep^-_{kR}$   \vphantom{$\overline{\slep_j}$}
     & $\nu_i \ell^-_j$, $\nu_j \ell^-_i$ 
     & $-$
     & $\ell^-_k \chio$
     & $-$ \\ \hline

     $\snu_{i}$, $\snu_{j}$  \vphantom{$\overline{\slep_j}$}     
     & $\ell^-_j \ell^+_k$, $\ell^-_i \ell^+_k$
     & $-$
     & $\nu_i \chio$, $\nu_j \chio$
     &  \\  \hline

     $ \tilde{\mathrm u}_{iR}$  \vphantom{$\overline{\slep_j}$}
     & $-$
     & $\bar{\mathrm d}_{j}\bar{\mathrm d}_{k}$
     & $\mathrm u_i \chio$
     & $-$ \\ \hline

     $ \tilde{\mathrm d}_{jR}, 
          \tilde{\mathrm d}_{kR} $   \vphantom{$\overline{\slep_j}$}
     & $-$
     & $\bar{\mathrm u}_{i}\bar{\mathrm d}_{k}$, 
        $\bar{\mathrm u}_{i}\bar{\mathrm d}_{j}$
     & $\mathrm d_j \chio, \mathrm d_k \chio$
     & $-$ \\ \hline

  \end{tabular}
  \icaption{R-parity violating decays of the supersymmetric
   particles considered in this analysis.
   Charged conjugate states are implied. Indirect decays
   via scalar leptons are relevant only for neutralinos when the scalar
   lepton is the LSP. Only 
   supersymmetric partners of the right-handed charged leptons are taken
   into account. Decays to more than three fermions are not listed.
   $\Zstar$ and $\Wstar$ indicate virtual Z and W bosons.
  \label{tab:decays}}
  \end{center}
\end{table*}

\section{Data and Monte Carlo Samples}

The data used  correspond to an integrated luminosity of 450.6 \pbi{}
collected with the L3 detector~\cite{L3}
at $\sqrt s = 192-208$ \GeV{}.
For the search for scalar quarks and scalar neutrinos decaying
via $\lambda''_{ijk}$ couplings, also the data sample collected 
at $\rts = 189$ \GeV{} is used. This corresponds to an additional integrated 
luminosity of 176.4 \pbi. 

The signal events are generated with the program {\tt SUSYGEN}~\cite{susygen}
for different mass values and
for all possible choices of the generation indices.

The following Monte Carlo
generators are used to simulate Standard Model background processes: 
{\tt PYTHIA}~\cite{pythia} for  
$ \mathrm \epem \ra \mathrm Z \, \epem$ and $\mathrm \epem \ra$ ZZ,
{\tt BHWIDE}~\cite{bhwide} for $\mathrm \epem \ra \epem $,
{\tt KK2F}~\cite{kk2f} for $\mathrm \epem \ra \mpmm$,
$\mathrm \epem \ra \tautau$ and $\mathrm \epem \ra \qqbar$,
{\tt PHOJET}~\cite{phojet} and {\tt PYTHIA}  
for $\mathrm \epem \ra \epem$ hadrons,
{\tt DIAG36}~\cite{diag36} for $\mathrm \epem \ra 
\epem \ell^+ \ell^-$ ($\mathrm{\ell = e, \mu, \tau}$),  
{\tt KORALW}~\cite{koralw} for $\mathrm \epem \ra \WpWm $ and
{\tt EXCALIBUR}~\cite{EXCALIBUR} for
$\ee \rightarrow \mathrm{\qqbar' \, \ell \nu}$ and 
$\epem \ra \ell\nu\ell '\nu$.
The number of simulated events corresponds to 
at least 50 times the luminosity of the data,
except for Bhabha and two-photon processes, 
where the Monte Carlo samples correspond to 
2 to 10 times the luminosity.

The detector response is simulated using the {\tt GEANT} 
package~\cite{geant}. It takes into account effects of energy loss,
multiple scattering and showering in the detector materials. Hadronic 
interactions are simulated with the
{\tt GHEISHA} program~\cite{gheisha}. Time dependent detector inefficiencies
are also taken into account in the simulation procedure.

Data and Monte Carlo samples are reconstructed with the same program.
Isolated leptons ($\ell = $ e, $\mu, \tau$) are identified as 
described in Reference~\citen{rpv2_l3}.
Remaining clusters and tracks are classified as hadrons. Jets are 
reconstructed with the DURHAM algorithm~\cite{durham}.
The jet resolution parameter $y_{mn}$ is defined
as the $\ycut$ value at which the event configuration changes from $n$ to 
$m$ jets. 
At least one time of flight measurement 
has to be consistent with the beam crossing to reject cosmic rays.

\section{\boldmath{$\lambda_{ ijk}$} Analysis}
\label{par:lambda_anal}

The different topologies arising when $\lambda_{ijk}$ couplings
dominate are shown in Table~\ref{tab:topologies} and 
can be classified into four categories: 
$2 \ell + \Emiss$, $4 \ell + \Emiss$, $6 \ell$,
$\ge 4 \, \ell$ plus possible jets and $\Emiss$.
The missing energy $\Emiss$ 
indicates final state neutrinos escaping detection.
After a common preselection~\cite{rpv2_l3}, based on the visible energy,
the event multiplicity and the number of identified leptons,
a dedicated selection is developed for each group, taking 
into account lepton flavours, particle boosts and 
virtual W and Z decay products. 

\begin{table*} [htbp] 
  \begin{center}
  \begin{tabular}{|l|l|l|} \hline 

     \multicolumn{1}{|l} {Direct decays}
    &\multicolumn{1}{l|} { }
    & Selections \\ \hline

     \multicolumn{1}{|l}   {\epem\ra~$\chim\chin \ra$} \rpvpha
    &\multicolumn{1}{l|}  {\hspace{-0.3 cm}{\em $\ell\ell\ell\ell$}$\nu\nu$ } 
    &   4 $\mathrm{\ell}$
      + $\Emiss$                  \\ \hline

   \multicolumn{1}{|l}  { \epem\ra~$\chap\cham \ra$} \rpvpha
   &\multicolumn{1}{l|} {\em\hspace{-0.3 cm}$\ell\ell\ell\ell\ell\ell$}
   &   6   $\mathrm{\ell}$   \\

   \multicolumn{1}{|l}   {} 
   &\multicolumn{1}{l|} {{\em\hspace{-0.3 cm}$\ell\ell\ell\ell$}$\nu\nu$} 
   &   4  $\mathrm{\ell}$ + $ \Emiss$ \\

   \multicolumn{1}{|l}   {} 
   &\multicolumn{1}{l|} {{\em\hspace{-0.3 cm}$\ell\ell$}$\nu\nu\nu\nu$ } 
   &   2  $\mathrm{\ell}$ + $ \Emiss$  \\ \hline 

    \multicolumn{1}{|l}   { \epem\ra~$\slepp\slepm \;\ra$} \rpvpha
   &\multicolumn{1}{l|} {\hspace{-0.3 cm}$\ell\nu\ell\nu$}
   &   2  $\mathrm{\ell}$ + $ \Emiss$  \\ \hline 

    \multicolumn{1}{|l}   { \epem\ra~$\snu\snu \hspace{0.5 cm}\ra$} \rpvpha
   &\multicolumn{1}{l|} {\hspace{-0.3 cm}$\ell\ell\ell\ell$}
   &   4  $\mathrm{\ell}$  + $ \Emiss$ \\ \hline 

     \multicolumn{1}{|l} {Indirect decays}
    &\multicolumn{1}{l|} { }
    &  \\ \hline

     \multicolumn{1}{|l}  { \epem\ra~$\chim\chin_{(n\geq 2)}$} \rpvpha
    &\multicolumn{1}{l|} {\hspace{-0.4 cm} \ra $\;$ cascades }
    & $\ge 4 \, \mathrm{\ell}$ + (jets) + $ \Emiss$  \\ \hline

     \multicolumn{1}{|l} {\epem\ra~$\chap\cham \ra$ } \rpvpha
    &\multicolumn{1}{l|}  {\hspace{-0.5 cm}
        $\tilde{\chi}_{1(2)}^0\tilde{\chi}_{1(2)}^0 \Wstar\Wstar$}
    & $\ge 4 \, \mathrm{\ell}$ + (jets) + $ \Emiss$  \\ \hline

    \multicolumn{1}{|l}   { \epem\ra~$\slepp\slepm \;\ra$} \rpvpha
   &\multicolumn{1}{l|} {\hspace{-0.3 cm}$\ell\ell\ell\ell\ell\ell\nu\nu$}
   & $\ge 4 \, \mathrm{\ell}$ + (jets) + $ \Emiss$  \\ \hline

    \multicolumn{1}{|l}   { \epem\ra~$\snu\snu \hspace{0.5 cm}\ra$} \rpvpha
   &\multicolumn{1}{l|} {\hspace{-0.3 cm}$\ell\ell\ell\ell\nu\nu\nu\nu$}
   &   4  $\mathrm{\ell}$  + $\Emiss$  \\ \hline

  \end{tabular}
  \icaption{Processes considered in the 
    $\mathrm \lambda_{ijk}$  analysis
    and corresponding selections~\protect\cite{rpv2_l3}. 
    $\chim\chin$ indicates neutralino pair-production 
    with $m = 1,2 \,$ and $n = 1,..,4$. ``Cascades''  refers
    to all possible final state combinations of Table~\ref{tab:decays}.
  \label{tab:topologies}}
  \end{center}
\end{table*}

After the preselection is applied, 
2567 events are selected in the data sample and $2593 \pm 12$ events
are expected from Standard Model processes. The main contributions are:
44.5\% from $\mathrm \WpWm$, 21.5\% from $\qqbar$,
14.7\% from $\mathrm{\qqbar'\, e \nu}$, 6.6\% from
two-photon processes
(3.9\% from $\epem \ell^+ \ell^-$ and 2.7\% from $\epem$\,hadrons),
 and 5.6\% from $\tautau$ events.

\begin{figure}[htbp]
\begin{center}
    \includegraphics[width=17cm, height=18cm]{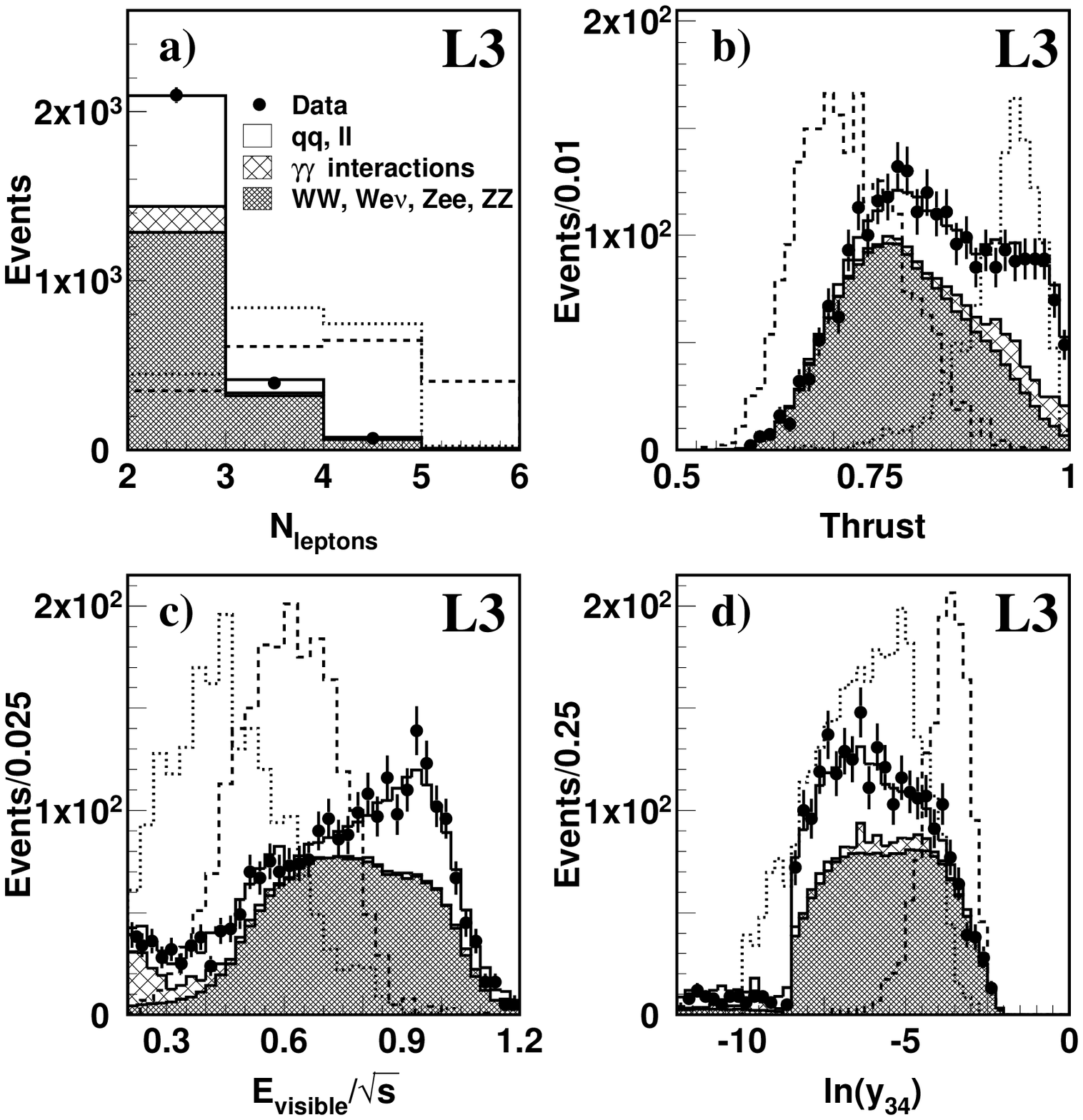}
  \end{center}
\vspace{-1.0truecm}
  \icaption{Data and Monte Carlo
 distributions of a) the number of leptons, b) thrust, 
 c) the normalised visible
 energy and  d) ln(${y_{34}}$) after the 
 $\mathrm \lambda_{ijk}$ preselection.
 The solid histograms show the expectations for Standard Model 
 processes.
The dotted and dashed histograms
 show two examples of signal, with dominant coupling $\lambda_{133}$.
The dotted histograms represent the process $\mathrm \epem \ra 
 \chio\chio$, for $\mchi =42$ \GeV,
corresponding to two hundred times the luminosity of the data.
The dashed ones
represent $\mathrm \epem \ra \chap\cham$, with $\mcha = 92$ \GeV{} and
 $\DM = \mcha - \mchi = 50$ \GeV{},
corresponding to twenty times this luminosity.
  \label{fig:ps_lambda}}
\end{figure}

Figure \ref{fig:ps_lambda} shows the distributions of the
number of leptons, thrust, normalised visible
energy and ln(${y_{34}}$) after the
preselection.
The data are in good agreement with the Monte Carlo expectations.

The final selection criteria are discussed in
Reference~\citen{rpv2_l3} and yield
the efficiencies for direct and indirect decays of the supersymmetric
particles summarized in Tables~\ref{tab:efficiencies1} 
and~\ref{tab:efficiencies2}, respectively.
Here and in the following sections we discuss only the results obtained 
for those choices of the generation indices which give the
lowest selection efficiencies. The quoted results
will thus be conservatively valid for any ${ijk}$ combination.
In the case of direct R-parity violating decays, the efficiencies 
are estimated for different mass values of the pair-produced
supersymmetric particles. In the case of indirect decays,
the efficiencies are estimated for different masses and $\DM$ ranges.
$\DM$ is defined as the mass difference $M_{susy} - \mchi$, where
$M_{susy}$ is the mass of the 
supersymmetric particle under investigation.

For direct neutralino or chargino decays,
as well as for all indirect decays studied,
the lowest efficiencies are found for $\lambda_{ijk} = \lamtre$,
due to the presence in the final state  of taus, whose detection
is more difficult.

\begin{table*} [htbp]
  \begin{center}
  \begin{tabular}{|l|l|l|c|c|c|c|c|c|c|c|} \hline
     \multicolumn{11}{|c|} {Direct decays } \\ \hline
     Coupling   & Process  & $M$ & 30 & 40 & 50 & 60 & 70 & 80 & 90 & 102 \\ \hline 
 $ \lambda_{133}$   & $\chim\chin$ \rpvpha 
         & $\epsilon$ & 15  & 24 & 32 & 37 & 40 & 42 & 45 & 46 \\ \cline{3-11}
    &    & $\sigma$   &  0.07 & 0.05 & 0.04 & 0.03 & 0.03  & 0.03   & 0.02  & 0.07   \\ \hline
 $ \lambda_{133}$   &  $\chap\cham$ \rpvpha 
         & $\epsilon$ &  --   &  -- & --  & --  & 38 & 40 & 43 & 43 \\ \cline{3-11}
    &    & $\sigma$  & -- &  -- & --  & --  & 0.07   & 0.06   & 0.06  & 0.17   \\ \hline
    $ \lambda_{12k} $   & $\slepp\slepm$ \rpvpha 
         & $\epsilon$ &  --   &  -- & --  & --  & 6 & 6  &8 & 6 \\ \cline{3-11}
    &    & $\sigma$  & -- &  -- & --  & --  & 0.39 & 0.36  & 0.27  & 1.16   \\ \hline
    $ \lambda_{121}$    & $\snu\snu$  \rpvpha   
         & $\epsilon$ &  --   &  -- & --  & --  &  6 &  8 & 7   &  5   \\ \cline{3-11}
    &    & $\sigma$   & -- &  -- & --  & --  & 0.20  & 0.15  & 0.17 & 0.68   \\ \hline
 $ \lambda''_{212}$ & $\chim\chin$, $\chap\cham$ \rpvpha
         & $\epsilon$ & 39  & 49 & 40 & 44 &42 & 43 &46 & 56 \\ \cline{3-11}
    &    & $\sigma$   & 0.11 & 0.10 & 0.08 & 0.12 & 0.12 & 0.11 & 0.10 & 0.18    \\ \hline
     $ \lambda''_{212}$   & $\serp\serm$ *, $\smurp\smurm$ *    \rpvpha
         & $\epsilon$ & 39  & 49 & 40 & 44 &42 & 43 &46 & 56 \\ \cline{3-11}
    &    & $\sigma$   & 0.11 & 0.10 & 0.08 & 0.12 & 0.12 & 0.11 & 0.10 & 0.18    \\ \hline
     $ \lambda''_{212}$   & $\staurp\staurm\,$ * \rpvpha
         & $\epsilon$ & 39 & 49 & 38 & 44 & 42 & 19 &14 & 13 \\ \cline{3-11}
    &    & $\sigma$  & 0.11  & 0.10 & 0.15 & 0.12 & 0.11  & 0.18 & 0.22 & 0.28 \\ \hline
     $ \lambda''_{212}$   & $\snu\snu$ * \rpvpha
         & $\epsilon$ & 7  & 14 & 29 & 21 & 21 & 22 &25 & 56 \\ \cline{3-11}
    &    & $\sigma$  & 0.66  & 0.16 & 0.13 & 0.18 & 0.18  & 0.17 & 0.15 & 0.18 \\ \hline
     $ \lambda''_{212}$   & $\sq\sq$  \rpvpha
         & $\epsilon$ & 27  & 26 & 22 & 32 & 31 & 34 &34 & 34 \\ \cline{3-11}
    &    & $\sigma$  & 0.10  & 0.13 & 0.07 & 0.05 & 0.28  & 0.27 & 0.16 & 0.13 \\ \hline

  \end{tabular}
  \icaption{Efficiency values ($\epsilon$, in \%) and 95\% C.L. 
   cross section upper limits ($\sigma$, in pb) for 
    direct decays of the supersymmetric particles, as a function of their
    mass ($M$, in \gev). As an example the
    efficiencies at $\rts = 206$ \GeV{} are shown, for the most conservative
    choice of the couplings. At the other centre-of-mass
    energies they are compatible within the uncertainties. Typical
    relative errors on the signal efficiencies, due to Monte Carlo
    statistics, are between 2\% and 5\%.      
    $\chim\chin$ indicates neutralino pair-production 
    with $m = 1,2 \,$ and $n = 1,..,4$.
    For direct neutralino decays we quote the $\chio\chio$ efficiencies. 
    The upper limits on the pair-production cross sections are calculated
    using the full data sample, with a total luminosity of 627 $\pbi$,
    except for the last mass point, where only the data collected at
    $\rts \ge 204$ \GeV{} are used, corresponding to a luminosity of 216 \pbi.
    Chargino and scalar lepton pair-production via $\lambda_{ijk}$
    couplings are not investigated  for mass values 
    excluded in Reference~\citen{rpv2_l3}.    
    For the processes marked with~* we refer to four-body decays, as 
    described in Section~\ref{par:lambdasec_anal}.
      \label{tab:efficiencies1}}
  \end{center}
\end{table*}

\begin{table*} [htbp]
  \begin{center}
  \begin{tabular}{|l|l|l|c|c|c|c|c|c|c|c|c|c|} \hline
     \multicolumn{13}{|c|} {Indirect decays }\\ \hline
    Coupling & Process & $\DM$ & 10 & 20 & 30 & 40 & 50 & 60 & 70 & 80 & 90 & 100  \\ \hline 
  $ \lambda_{133}$   & $\chim\chin_{(n\geq 2)}$  \rpvpha
      & $\epsilon$ & 49 & 48 & 48 & 47 & 45 & 43 & 41 & 38 &36 & 35\\ \cline{3-13}
  &   & $\sigma$  & 0.09 & 0.09  & 0.09 & 0.09 & 0.10 & 0.10 & 0.11 & 0.12 & 0.12 & 0.13 \\ \hline

$ \lambda_{133}$   & $\chap\cham$ \rpvpha 
         & $\epsilon$ & 47 & 43 & 39 & 34 & 31 & 25 & 20 & -- & -- & -- \\ \cline{3-13}
    &    & $\sigma$ & 0.08 & 0.09 & 0.10 & 0.11 & 0.12 & 0.15 & 0.18 & -- & -- & -- \\ \hline
 $ \lambda_{133}$   & $\serp\serm$   \rpvpha
         & $\epsilon$ & 61 & 62 & 63 & 54 & 46 & 35 & 24 & -- & -- & -- \\ \cline{3-13}
    &    & $\sigma$  & 0.06  & 0.06  & 0.06 & 0.07 & 0.08 & 0.11 & 0.15 & -- & -- & -- \\ \hline
 $ \lambda_{133}$   & $\smurp\smurm$  \rpvpha
         & $\epsilon$ & 71 & 76 & 80 & 77 & 75 & 70 & 65 & -- & -- & -- \\ \cline{3-13}
    &    & $\sigma$  & 0.05 & 0.05 & 0.05 & 0.05   & 0.05  & 0.05   & 0.06  & --  & -- & --  \\ \hline
 $ \lambda_{133}$   & $\staurp\staurm$  \rpvpha
         & $\epsilon$ & 52 & 59 & 66 & 65 & 64 & 60 & 56 &  -- & -- & -- \\ \cline{3-13}
    &    & $\sigma$ & 0.07 & 0.06 & 0.06 & 0.06 & 0.06  & 0.06 & 0.07  & --  & -- & --  \\ \hline
 $ \lambda_{133}$   & $\snu\snu$    \rpvpha 
         & $\epsilon$ & 50 & 49 & 49 & 43 & 41 & 39 & 36 & -- & -- & -- \\ \cline{3-13}
    &    & $\sigma$  & 0.07 & 0.07 & 0.07 & 0.08 & 0.08 & 0.09 & 0.10 & -- & -- & -- \\ \hline
$ \lambda''_{212}$ & $\chim\chin_{(n\geq 2)}$   \rpvpha
         & $\epsilon$ & 57 & 60 & 63 &68 & 66 & 64 & 62 &  58 & 54 & 46 \\ \cline{3-13}
    &    & $\sigma$ & 0.18 & 0.17 & 0.16& 0.15 & 0.15 &0.16 &0.17 & 0.18 & 0.20 & 0.23 \\ \hline
$ \lambda''_{212}$ & $\chap\cham$ \rpvpha 
         & $\epsilon$ & 65 & 70 & 69 & 73 & 72 & 70 & 71 & -- & -- & -- \\ \cline{3-13}
    &    & $\sigma$  & 0.16 & 0.15 & 0.15 & 0.14 & 0.15 & 0.15 & 0.15 & --  & --  & --  \\ \hline
 $ \lambda''_{212}$   & $\serp\serm$  \rpvpha 
         & $\epsilon$ & 29 & 51 & 56 & 63 & 66 & 69 & 56 &  46 & 36 & -- \\ \cline{3-13}
    &    & $\sigma$  & 0.18 & 0.09 & 0.05 & 0.05 & 0.05 & 0.05 & 0.05 & 0.06 & 0.08 & --  \\ \hline
 $ \lambda''_{212}$   & $\smurp\smurm$   \rpvpha
         & $\epsilon$ & 20 & 28 & 41 & 49 & 52 & 55 & 52 &  42& 27 & -- \\ \cline{3-13}
    &    & $\sigma$  & 0.10 & 0.05 & 0.05  & 0.05 & 0.05 & 0.05 & 0.05 & 0.06 & 0.09 & --  \\ \hline
 $ \lambda''_{212}$   & $\staurp\staurm$ \rpvpha
         & $\epsilon$ & 53 & 57 & 63 & 56 & 46 & 40 & 29 &  17 & 13 & -- \\ \cline{3-13}
    &    & $\sigma$  & 0.15 & 0.13 & 0.13 & 0.13 & 0.15 & 0.16 & 0.22 & 0.23 & 0.24 & -- \\ \hline
 $ \lambda''_{212}$   & $\snu\snu$    \rpvpha 
         & $\epsilon$ & 41 & 43 & 44 & 39 & 37 & 32 & 40 &  50 & 35 & -- \\ \cline{3-13}
    &    & $\sigma$ & 0.13 & 0.12 & 0.12 & 0.12 & 0.14 & 0.15 & 0.08 & 0.11 & 0.12 & -- \\ \hline
 $ \lambda''_{212}$   & $\sq\sq$    \rpvpha 
         & $\epsilon$ & 55 & 59 & 64 & 65 & 63 & 58 & 47 &  45 & 43 & -- \\ \cline{3-13}
    &    & $\sigma$  & 0.18 & 0.16 & 0.15 & 0.15 & 0.16 & 0.17 & 0.22 & 0.22 & 0.23 & -- \\ \hline

  \end{tabular}
  \icaption{Efficiency values ($\epsilon$, in \%) and 95\% C.L. 
   cross section upper limits ($\sigma$, in pb) for 
    indirect decays of the supersymmetric particles, as a function of 
    $\DM$  (in \gev). As an example the
    efficiencies at $\rts = 206$ \GeV{} are shown, for the most conservative
    choice of the couplings. At the other centre-of-mass
    energies they are compatible within the uncertainties. Typical
    relative errors on the signal efficiencies, due to Monte Carlo
    statistics, are between 2\% and 5\%.      $\chim\chin$ 
    indicates neutralino pair-production 
    with $m = 1,2 \,$ and $n = 2,..,4$. The efficiencies correspond to
    $ \mchim + \mchin = $ 206 \GeV.
    For indirect decays of charginos, scalar leptons and scalar quarks,
    the selection efficiencies correspond to a mass of 102 \GeV.
    The upper limits on the pair-production cross sections are calculated
    using the data at $\rts \ge 204 \GeV{}$, with an integrated luminosity
    of 216 \pbi.
      \label{tab:efficiencies2}}
  \end{center}
\end{table*}

In the case of pair-production of scalar charged leptons, followed
by direct decays via $\lambda_{ijk}$, the final state contains 
two leptons plus missing energy. The lepton flavours are given
by the indices $i$ and $j$, independently of the value of $k$.
The lowest selection efficiency is found for $\lambda_{ijk} = \lambda_{12k}$,
{\it i.e.} for events with electrons and muons in the final state,
since these low multiplicity events require a tight
selection to suppress the large background from
lepton pair-production.
 
Direct decays of scalar neutrinos yield four leptons in the final state. 
The $4 \ell + \Emiss$ selections are used as they provide
a good analysis sensitivity
comparable to that of the dedicated selections
for scalar electrons, muons and taus.
Scalar neutrino decays into electrons 
and muons are selected with lower efficiency than decays
into taus, due to the missing energy requirements.
In particular, the lowest efficiency is obtained for
$\lambda_{121}$, which can give rise to the decays 
$\mathrm{\snu_e \ra \mu^- e^+\,}$ and  $\mathrm{\snu_\mu \ra e^- e^+}$.

\section{\boldmath{$\lambda''_{ ijk}$} Analysis}
\label{par:lambdasec_anal}

When the $\lambda''_{ijk}$ couplings dominate,
the flavour composition depends on the generation indices.
In the case of neutralino and chargino pair-production, 
the different topologies can be classified into two groups:
multijets and multijets with leptons and/or missing energy,
as shown in Table~\ref{tab:topologies_udd}.
After a common preselection~\cite{rpv2_l3}, dedicated selections are
developed for each group, depending on the particle boosts,
the $\DM$ values and the 
virtual W decay products.

\begin{table*} [htbp] 
  \begin{center}
  \begin{tabular}{|l|l|l|} \hline 

     \multicolumn{1}{|l} {Direct decays}
    &\multicolumn{1}{l|} { }
    & Selections \\ \hline

     \multicolumn{1}{|l}   {\epem\ra~$\chim\chin \;\ra$} \rpvpha
    &\multicolumn{1}{l|}  {\hspace{-0.3 cm}{\rm qqqqqq}}
    &    multijets                  \\ \hline

   \multicolumn{1}{|l}  { \epem\ra~$\chap\cham \ra$} \rpvpha
   &\multicolumn{1}{l|}   {\hspace{-0.3 cm}{\rm qqqqqq}}
   &   multijets  \\   \hline

     \multicolumn{1}{|l} {Indirect decays}
    &\multicolumn{1}{l|} { }
    &  \\ \hline

     \multicolumn{1}{|l}  { \epem\ra~$\chim\chin_{(n\geq 2)} \ra$} \rpvpha 
    &\multicolumn{1}{l|}  {\hspace{-0.3 cm}{\rm qqqqqq qq}}
    & multijets   \\ 

     \multicolumn{1}{|l}  { } 
    &\multicolumn{1}{l|}   {\hspace{-0.3 cm}{\rm qqqqqq $\ell\ell$}}
    & multijets + lepton(s)  \\ 

     \multicolumn{1}{|l}  { } 
    &\multicolumn{1}{l|}   {\hspace{-0.3 cm}{\rm  qqqqqq $\nu\nu$}}
    & multijets  \\ \hline

     \multicolumn{1}{|l} {\epem\ra~$\chap\cham \ra$ } \rpvpha
    &\multicolumn{1}{l|}  {\hspace{-0.5 cm}{qqqqqq  qqqq}}
    & multijets \\

     \multicolumn{1}{|l} { } 
    &\multicolumn{1}{l|}   {\hspace{-0.5 cm}{qqqqqq  qq $\ell\nu$}}
    & multijets + lepton(s) \\

     \multicolumn{1}{|l} { } 
    &\multicolumn{1}{l|}   {\hspace{-0.5 cm}{qqqqqq $\ell\ell\nu\nu$}}
    & multijets + lepton(s)  \\ \hline

    \multicolumn{1}{|l}   { \epem\ra~$\slepp\slepm \;\ra$} \rpvpha
   &\multicolumn{1}{l|} {\hspace{-0.6 cm} qqqqqq $\ell\ell $}
   &   6 jets + 2 $\mathrm{\ell}$   \\ \hline 

    \multicolumn{1}{|l}   { \epem\ra~$\snu\snu \hspace{0.5 cm}\ra$} \rpvpha
   &\multicolumn{1}{l|} {\hspace{-0.6 cm} qqqqqq $\nu\nu$}
   &   6 jets + $\Emiss$  \\ \hline 

    \multicolumn{1}{|l}   { \epem\ra~$\sq\sq \hspace{0.5 cm}\ra$} \rpvpha
   &\multicolumn{1}{l|} {\hspace{-0.6 cm} qqqqqq qq}
   &   multijets  \\ \hline

  \end{tabular}
  \icaption{Processes considered in the 
    $\mathrm \lambda''_{ijk}$ analysis and corresponding
    selections~\protect\cite{rpv2_l3}. For masses below 50~\gev{}
    or small $\DM$ values not all jets in the event can be resolved.
    $\chim\chin$ indicates neutralino pair-production 
    with $m = 1,2 \,$ and $n = 1,..,4$.
    For final states with neutrinos we use selections with no explicit
    missing energy requirement, because for those topologies $\Emiss$
    is small, except for the scalar neutrino dacays.
  \label{tab:topologies_udd}}
  \end{center}
\end{table*}

\begin{figure}[htbp]
\begin{center}
    \includegraphics[width=17cm, height=18cm]{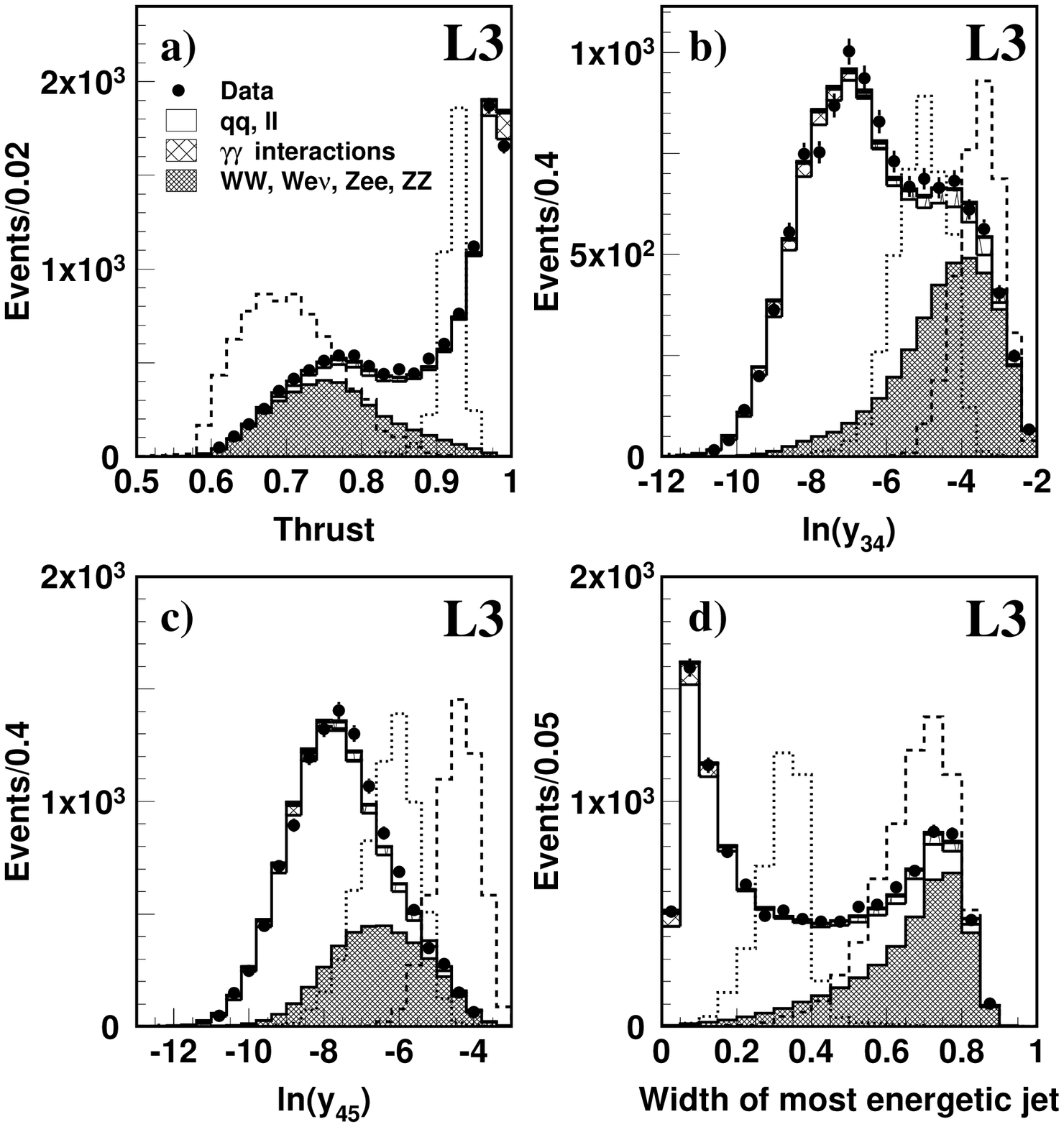}
  \end{center}
  \icaption{Data and Monte Carlo
distributions of a) thrust, b) ln(${y_{34}}$), c) ln(${y_{45}}$) 
 and d) width of the most energetic jet after the
 $\lambda''_{ijk}$ preselection. 
 The solid histograms show the expectations for Standard Model 
 processes.
 The dashed and dotted histograms show two examples of 
 signal, with dominant coupling $ \lamuno$, corresponding to decays into
c, d and s quarks.
The dotted histograms represent the process
 $\mathrm \epem \ra \chio\chio$, with $ \mchi = 40$ \GeV{},
corresponding to one hundred times the luminosity of the data. 
The dashed ones
represent $\mathrm \epem \ra \chap\cham$, with $\mcha = 90$ \GeV{} and
 $\DM = \mcha - \mchi = 60$ \GeV{},
corresponding to fifteen times this luminosity.
  \label{fig:ps_lambdasec}}
\end{figure}

In the case of neutralino, chargino, scalar charged lepton and
scalar quark pair-production, the preselection 
aims at selecting well balanced hadronic events and yields
11770 events in the data sample to be compared with $ 11719 \pm 31 $
expected from Standard Model processes, 
of which 62.0\%  are from $\mathrm \qqbar$ and 32.8\% $\mathrm \WpWm$.
Figure~\ref{fig:ps_lambdasec} shows the distributions of
thrust, ln($y_{34}$),  ln($y_{45}$) 
and width of the most energetic jet 
after the preselection.
The width of a jet is defined as $ {p}_{T}^{jet} / {E^{jet}_{ }}$,
where the event is clustered into exactly two jets,
and ${p}_{T}^{jet}$ is the sum of the projections of the particle 
momenta
on to a plane perpendicular to the jet axis, and ${E^{jet}_{ }}$
is the jet energy.
There is good agreement between data and Monte Carlo expectations.
The efficiencies for direct and indirect decays of the supersymmetric
particles after the selections discussed in Reference~\citen{rpv2_l3}
are summarized in Tables~\ref{tab:efficiencies1} 
and~\ref{tab:efficiencies2}, respectively.

Scalar quarks and scalar neutrinos,
not studied in our previous papers, are searched for as follows.
Scalar quark pairs can decay directly into 4 or indirectly into 8 quarks, 
as shown in Table~\ref{tab:decays}.
In the first case, the main background
sources are $\qqbar$ events and $\WpWm$ decays. For low masses
of the primary scalar quarks, the signal configuration is more similar
to two back-to-back jets, due to the large jet boost. In this case
we use the least energetic jet width to reject the $\qqbar$ background,
which is the dominant one at low masses.
For larger scalar quark masses ($M_{\tilde{\mathrm q}} > 50$ \GeV), 
the signal events are better described by a 4-jet configuration and
selection criteria are applied on $y_{34}$ and the
$\chi^2$ of a kinematical fit, which imposes four-momentum conservation
and equal mass constraints.
In the case of indirect decays into 8 quarks, the same selections
as for $\chio\chio$ decays into 6 quarks are used~\cite{rpv2_l3}.

For scalar neutrino pair-production, a different preselection is
performed, to take into account the missing momentum in the
final state. 
Low multiplicity events, such as leptonic Z and W decays, are
rejected by requiring at least 13 calorimetric clusters.
At least one charged track has to be present.
The visible energy has to be greater than $0.2\sqrt s$. 
In order to remove background contributions from two-photon interactions,
the energy in a cone of $\mathrm 12^\circ$ half-opening angle around the
beam axis has to be below 20\% of the total visible energy. 
Furthermore, the thrust axis is required to be well contained in the
detector. Unbalanced events
with an initial state radiation 
photon in the beam pipe are removed.
Semileptonic $\WpWm$ decays are rejected by the requirement
that neither the di-jet invariant mass nor that of any identified
lepton and the missing four-momentum 
should be in a 5 \GeV{} interval around the W mass.
This preselection yields 13950 events in the data at $\rts = 189-208$ \GeV{}
where $13662 \pm 45$ are expected from Standard Model processes and 
the main contributions are 50.6\% from $\qqbar$,
32.8\% from  $\WpWm$,  9.2\% from  $\epem\qqbar$ and
4.0\% from $\qqbar'$e$\nu$.
The difference in the number of found and expected data
appears in the region where the 
visible energy is below $0.5\sqrt s$, where an important contribution
from two-photon interactions and $\ell\nu\ell'\nu$ events is expected.
Such events are afterwards rejected by the optimization procedure,
which requires a high visible energy.

In the case of indirect decays of scalar neutrinos,
the only visible decay products are the jets coming from neutralino decays.
Therefore we have derived five selections according to the 
neutralino mass value,
reflecting the different boost and jet broadening configurations.
The final selection criteria are optimized~\cite{rpv2_l3} by
taking into account the 
following variables: jet widths,
ln$(y_{34})$ and ln$(y_{45})$,
visible energy and polar angles of the missing 
momentum vector and of the thrust axis.

Supersymmetric partners of the right-handed leptons have no direct
two-body decays via $\lambda''_{ijk}$ couplings.
However, when scalar leptons are lighter than $\chio$, 
the four-body decay $\slepr \ra \ell$qqq can occur~\cite{superp2} providing
 the same final state as that resulting from indirect
decays, but with virtual $\chio$ production.
The non-resonant four-body decay is not implemented in the 
generator.
For this reason, we use the
results of the indirect decay analysis, performing a scan
over all neutralino mass values up to
$M_{\slepr}$.
The resulting lowest efficiency is conservatively quoted 
in the following for 
four-body decays. It is found in most cases for $\mchi \simeq M_{\slepr}$,
as the resulting low energy lepton can not be resolved from the nearby
jet.
For scalar taus with masses above 70~\gev{},
the lowest efficiency is found for high $\DM$ values, as in the case
of indirect decays.

\section{Model Independent Results}
\label{par:mod_ind_results}

Table~\ref{tab:tab8}  shows the overall numbers of candidates and expected background
events for the different processes.  No significant excess of events is
observed.  Therefore upper limits are set on the neutralino, chargino and
scalar lepton pair-production cross sections assuming direct or indirect
R-parity violating decays.

In the case of $\lambda_{ijk}$ couplings, upper limits are set for each
process, independently of the mass value of the supersymmetric particle
considered. For $\lambda''_{ijk}$ couplings, upper limits are derived for
each process depending on the mass range of the supersymmetric particles,
since this procedure improves the sensitivity of analyses with high
background level.

These limits take into account the estimated
background contamination.
Systematic uncertainties on the signal efficiency 
are dominated by Monte Carlo statistics.
The typical relative error is between 2\% and 5\%
 and it is included in the
calculations of the signal upper limits~\cite{upperlimit}.

Tables~\ref{tab:efficiencies1} and~\ref{tab:efficiencies2}
show the 95\% confidence level (C.L.) upper limits on 
supersymmetric particle pair-production cross~sections. 
For each mass point, all data collected at
centre-of-mass energies above the production threshold
are combined. For low mass values, the data
at $\rts =$ 189 \GeV{} are also used.
Therefore these upper limits 
should be interpreted as a limit on the luminosity-weighted average 
cross section.

\begin{table*} [htbp]
  \begin{center}
  \begin{tabular}{|l|l|rcl|r|} \hline 
             Coupling &  Process  & \multicolumn{3}{|c|} {$N_{back}$} & $N_{data}$ \\ \hline
$ \lambda_{ijk}$  & $\chio\chio$ \rpvpha & \pho 4.9 &$\pm$& 0.5 \pho &  6   \\ \cline{2-6}
                  & $\chim\chin$  \rpvpha & \pho 14.7 &$\pm$& 0.6 \pho & 15  \\ \cline{2-6} 
      & $\chap\cham$ (indirect)  \rpvpha  & \pho10.1 &$\pm$& 0.3\pho & 10  \\ \cline{2-6}
        & $\chap\cham$ (direct) \rpvpha  & \pho37\phopunto &$\pm$& 3  \pho& 40   \\ \cline{2-6}
     & $\slepp\slepm$ (indirect) \rpvpha & \pho10.1 &$\pm$& 0.3\pho  & 10   \\ \cline{2-6}
       & $\slepp\slepm$ (direct) \rpvpha & \pho31\phopunto  &$\pm$& 2 \pho & 34   \\ \cline{2-6}
                  & $\snu\snu$  \rpvpha  & \pho4.9 &$\pm$& 0.5 \pho &  6   \\ \hline
$  \lambda''_{ijk}$ & $\chio\chio$ \rpvpha & \pho661 &$\pm$& \phouno4 \pho& 605  \\ \cline{2-6}
                    & $\chap\cham$ \rpvpha & \pho446 &$\pm$& \phouno3 \pho& 404   \\ \cline{2-6}
                    & $\slepp\slepm$\rpvpha& \pho413 &$\pm$& \phouno2 \pho& 361   \\ \cline{2-6}
                  & $\snu\snu$  \rpvpha  & \pho671 &$\pm$& \phouno6  \pho&  669   \\  \cline{2-6}
                  & $\sq\sq$  \rpvpha  & \pho3387 &$\pm$& 13  \pho&  3411  \\ \hline
  \end{tabular}
  \icaption{Number of observed data ($N_{data}$) and expected background 
  ($N_{back}$) events  for the different processes.
      The uncertainty on the
     expected background is due to Monte Carlo statistics. The deficit in the
number of observed data in the neutralino, chargino and slepton analyses is
correlated among the channels.
 \label{tab:tab8}}
  \end{center}
\end{table*}

\section{Interpretation in the MSSM}
\label{par:MSSM}

In the MSSM framework, neutralino and chargino masses, couplings and
cross sections depend on the gaugino mass parameter, ${M_2}$, 
the higgsino mass mixing parameter, $\mu $, the ratio
of the vacuum expectation values of the two Higgs doublets, $\tan\beta$, and
the common mass of the scalar particles at the GUT scale, ${m_0}$.
The results presented in this section are obtained by performing a scan 
in the ranges:
 $\,0 \leq M_2 \leq 1000$ \GeV,
$- 500$ \GeV{} $\leq \mu \leq$ 500 \GeV, $0 \leq m_0 \leq 500$ \gev{}
and $0.7 \leq \tan\beta \leq 40$.
They do not depend on the value of the
trilinear coupling in the Higgs sector, $A$.

\subsection{Mass Limits from Scalar Lepton and Scalar Quark Searches }
\label{par:mass_limits_2}

For scalar lepton and scalar quark pair-production,
mass limits are derived by direct comparison of  
the 95\% C.L. cross section upper limits
with the scalar particle
pair-production cross sections, which depend on the scalar
particle mass.

We assume no mixing in the scalar lepton sector.
Scalar electron and scalar electron neutrino
pair-production have an additional contribution from
the $t$-channel exchange of a neutralino or
chargino, whose mass spectrum depends on the MSSM parameters.
In this case the mass limits are
derived at a given value of \tb \, and $\mu$, here chosen  
to be $\tb = \sqrt{2}$ and $\mu = -200$ \GeV.
For scalar quarks, mixing is taken into account for
the third generation. The cross section depends on the
scalar quark mass and on the mixing angle $\theta_{LR}$. 
For  $\rts = 189-208$ \GeV{}
the production cross section for scalar top pairs 
is minimal for $\cos\theta_{LR} \sim 0.51$ and for scalar
bottom pairs for $\cos\theta_{LR} \sim 0.36$.
These values are
conservatively used in this analysis.

Figures~\ref{fig:fig3} and~\ref{fig:fig4} show the excluded
95\% C.L. contour for different scalar lepton
and scalar quark masses, as a function of the neutralino
mass. Indirect decays of the scalar leptons dominate 
over direct ones in the region with $\DM > 2 \,$\gev{}.
For $0 \le \DM < 2 \,$\gev, 100\% branching ratio
either into direct or indirect decays is assumed  
and the worst result is shown.
In the negative $\DM$ region only direct decays contribute.
For $\lambda''_{ijk}$ direct decays of the scalar leptons 
we quote the results from four-body processes.
The 95\% C.L. lower mass limits are shown in 
Table~\ref{tab:limits_tot2}, for both direct and indirect decays.

\begin{table*} [htbp]
  \begin{center}
  \begin{tabular}{|l|c|c|c|c|c|c|c|c|c|c|c|} \hline 
 
 Mass Limit (\gev)& $M_{\serr}$ & $M_{\smur}$ &
  $M_{\staur}$ &  $M_{\snumt}$   & $M_{\snue}$ &  
 $M_{\qur}$  &  $M_{\qul}$  &  $M_{\qdr}$  &
  $M_{\qdl}$  &   $M_{\stop}$ &   $M_{\sbottom}$ 
\vphantom{ $M^2_{\qdl}$}\\ \hline

 $\lambda_{ijk}$ (direct) & 69 & 61 & 61 & 65 & 95 &  --   & --   & --   & --   & --   
    & --    \\ \hline
    $\lambda_{ijk}$ (indirect) & 79 & 87 & 86 & 78 & 99 &  --   & --   & --   & --   & --   
    & --    \\ \hline

$\lambda''_{ijk}$ (direct) & 96 & 86 & 75 & 70 & 99 & 80 & 87  & 56 & 86 & 77 & 55 
\\ \hline
$\lambda''_{ijk}$ (indirect) & 96 & 86 & 75 & 70 & 99 & 79 & 87 & 55 & 86 & 77 & 48 
\\ \hline

\end{tabular}
 \icaption{Lower limits at 95\% C.L. on the masses of the scalar leptons
   and scalar quarks. The limits result from direct comparison of
   the 95\% C.L. cross section upper limits  with the scalar particle
   pair-production cross sections. $\qur$, $\qul$, $\qdr$ and $\qdl$
   refer to any type of up and down supersymmetric partners
 of the right-handed and left-handed quarks. $\stop$ and $\sbottom$
 limits are quoted in the case of minimal production cross section.
 For $\lambda''_{ijk}$ direct decays of
scalar leptons we refer to four-body processes.
 \label{tab:limits_tot2}}
\end{center}
\end{table*}

\subsection{Mass Limits from Combined Analyses }
\label{par:mass_limits_1}

A point in the MSSM parameter space is excluded if the total number
of expected events is greater than the combined upper limit at 95\% C.L.
on the number of signal events. Neutralino, chargino, scalar
lepton and scalar quark analyses are combined since several processes 
can occur at a given point.  Gaugino and scalar mass unification at the
GUT scale is assumed.
The constraints from the L3 lineshape measurements at the Z 
pole~\cite{gammalim}
are also taken into account~\cite{rpv2_l3}.
We derive lower limits at 95\% C.L. on the 
neutralino, chargino and scalar lepton masses, as detailed
in Table~\ref{tab:limits_tot}.

\begin{table*} [htbp]
  \begin{center}
  \begin{tabular}{|c|c|c|c|c|c|c|} \hline 
 
 Mass Limit (\gev)&   $\mchi$ &  $\mchid$ &   $\mchit$ &  $\mcha$ &
                  $M_{\slepr}$ &   $M_{\snu}$ \\ \hline

$\lambda_{ijk}$ & \phouno 40.2    & \phouno 84.0    &   107.2    & 
                 103.0    &\phouno 82.7   & 152.7 \\ \hline

$\lambda''_{ijk}$  &    \phouno 39.9  &    \phouno 80.0  &  107.2 &  
                   102.7  &  \phouno 88.7 &  149.0 \\ \hline     

\end{tabular}
 \icaption{Lower limits at 95\% C.L. on the masses of the supersymmetric
  particles considered in this analysis. The limits result from combined
 analysis at each MSSM point and from a global scan in the parameter
  space, as detailed in section~\ref{par:MSSM}.
The limits on $M_{\slepr}$ hold
  for ${\serr}$, $\smur$ and $\staur$.
 \label{tab:limits_tot}}
\end{center}
\end{table*}

Figure~\ref{fig:mlimit1} shows the 95\% C.L. lower limits on neutralino and
scalar lepton masses as a function of \tb. The $\chio$ and $\chid$ 
mass limits are shown for $m_0 = $ 500 \gev{} and the $\slepr$ ones for
$m_0 =$ 0. These values of $m_0$ correspond to the absolute minima from
the complete scan on $M_2$, $\mu$, ${m_0}$ and $\tan\beta$.
The chargino mass limit is almost independent of $\tan\beta$,
and is close to the kinematic limit for any value of $\tan\beta$ and $m_0$.
For high \mo values, neutralino and scalar lepton pair-production
contributions are suppressed and the mass limits 
are given mainly by the chargino exclusion.

For 0 $\leq{m_0}<$ 50 \GeV{} and 1 $\leq \tan\beta< 2$, the lightest
scalar lepton, the supersymmetric partner of the right-handed electron,
can be the LSP. Therefore in this region only 
the scalar lepton analysis contributes to the limit on the scalar lepton
mass. For higher values of \tb, $\chio$ is the LSP and the lower limit on the
scalar lepton mass is mainly given by the $\chio\chio$
exclusion contours. 
The absolute limit on $M_{\slepr}$ is found at $\tan\beta = 0.8$ in the
case of $\lambda_{ijk}$  and at $\tan\beta = 0.7$ for
$\lambda''_{ijk}$. The difference in the limits is due to the lower 
cross section upper limit
of $\lambda''_{ijk}$ for scalar lepton direct decays, since 
the limit on $M_{\slepr}$ is found when the $\slepr$
is the LSP.
The same limits are obtained without the assumption of a 
common scalar mass at the GUT scale.
For $\lambda_{ijk}$ the
bounds on the scalar lepton masses are found in the case in which
the $\slepr$  is the LSP. For $\lambda''_{ijk}$ the limits are found 
when  the $\slepr$ and $\chio$ are nearly degenerate in mass. In both
cases, the neutralino analyses give the main contribution to the exclusion
in the regions of the parameter space around the limit.
The remaining sensitivity is due to searches for direct slepton decays
via $\lambda_{ijk}$. As these searches are equally sensitive to scalar
electron, muon or tau signals, as shown in Table~\ref{tab:efficiencies1},
the limits are unchanged. 
The scalar neutrino mass limit is also mainly due to neutralino
exclusions, resulting in a 95\% C.L. lower limit on the scalar
neutrino mass above the kinematic limit.

The search for R-parity violating decays of supersymmetric particles 
reaches at least the same sensitivity as in the R-parity 
conserving case~\cite{rpc_susy2}.
Therefore, the supersymmetry limits obtained at LEP are independent of 
R-parity conservation assumptions.

%
\newpage
\typeout{   }     
\typeout{Using author list for paper 244 -- 246 }
\typeout{$Modified: Jul 31 2001 by smele $}
\typeout{!!!!  This should only be used with document option a4p!!!!}
\typeout{   }
%
%
%
%
%
%

\newcount\tutecount  \tutecount=0
\def\tutenum#1{\global\advance\tutecount by 1 \xdef#1{\the\tutecount}}
\def\tute#1{$^{#1}$}
\tutenum\aachen            
\tutenum\nikhef            
\tutenum\mich              
\tutenum\lapp              
\tutenum\basel             
\tutenum\lsu               
\tutenum\beijing           
\tutenum\berlin            
\tutenum\bologna           
\tutenum\tata              
\tutenum\ne                
\tutenum\bucharest         
\tutenum\budapest          
\tutenum\mit               
\tutenum\panjab            
\tutenum\debrecen          
\tutenum\florence          
\tutenum\cern              
\tutenum\wl                
\tutenum\geneva            
\tutenum\hefei             
\tutenum\lausanne          
\tutenum\lyon              
\tutenum\madrid            
\tutenum\florida           
\tutenum\milan             
\tutenum\moscow            
\tutenum\naples            
\tutenum\cyprus            
\tutenum\nymegen           
\tutenum\caltech           
\tutenum\perugia           
\tutenum\peters            
\tutenum\cmu               
\tutenum\potenza           
\tutenum\prince            
\tutenum\riverside         
\tutenum\rome              
\tutenum\salerno           
\tutenum\ucsd              
\tutenum\sofia             
\tutenum\korea             
\tutenum\utrecht           
\tutenum\purdue            
\tutenum\psinst            
\tutenum\zeuthen           
\tutenum\eth               
\tutenum\hamburg           
\tutenum\taiwan            
\tutenum\tsinghua          

{
\parskip=0pt
\noindent
{\bf The L3 Collaboration:}
\ifx\selectfont\undefined
 \baselineskip=10.8pt
 \baselineskip\baselinestretch\baselineskip
 \normalbaselineskip\baselineskip
 \ixpt
\else
 \fontsize{9}{10.8pt}\selectfont
\fi
\medskip
\tolerance=10000
\hbadness=5000
\raggedright
\hsize=162truemm\hoffset=0mm
\def\r{\rlap,}
\noindent

P.Achard\r\tute\geneva\ 
O.Adriani\r\tute{\florence}\ 
M.Aguilar-Benitez\r\tute\madrid\ 
J.Alcaraz\r\tute{\madrid,\cern}\ 
G.Alemanni\r\tute\lausanne\
J.Allaby\r\tute\cern\
A.Aloisio\r\tute\naples\ 
M.G.Alviggi\r\tute\naples\
H.Anderhub\r\tute\eth\ 
V.P.Andreev\r\tute{\lsu,\peters}\
% ...
\newpage
\begin{list}{A}{\itemsep=0pt plus 0pt minus 0pt\parsep=0pt plus 0pt minus 0pt
                \topsep=0pt plus 0pt minus 0pt}
\item[\aachen]
 I. Physikalisches Institut, RWTH, D-52056 Aachen, FRG$^{\S}$\\
 III. Physikalisches Institut, RWTH, D-52056 Aachen, FRG$^{\S}$
\item[\nikhef] National Institute for High Energy Physics, NIKHEF, 
     and University of Amsterdam, NL-1009 DB Amsterdam, The Netherlands
\item[\mich] University of Michigan, Ann Arbor, MI 48109, USA
\item[\lapp] Laboratoire d'Annecy-le-Vieux de Physique des Particules, 
     LAPP,IN2P3-CNRS, BP 110, F-74941 Annecy-le-Vieux CEDEX, France
\item[\basel] Institute of Physics, University of Basel, CH-4056 Basel,
     Switzerland
\item[\lsu] Louisiana State University, Baton Rouge, LA 70803, USA
\item[\beijing] Institute of High Energy Physics, IHEP, 
  100039 Beijing, China$^{\triangle}$ 
\item[\berlin] Humboldt University, D-10099 Berlin, FRG$^{\S}$
\item[\bologna] University of Bologna and INFN-Sezione di Bologna, 
     I-40126 Bologna, Italy
\item[\tata] Tata Institute of Fundamental Research, Mumbai (Bombay) 400 005, India
\item[\ne] Northeastern University, Boston, MA 02115, USA
\item[\bucharest] Institute of Atomic Physics and University of Bucharest,
     R-76900 Bucharest, Romania
\item[\budapest] Central Research Institute for Physics of the 
     Hungarian Academy of Sciences, H-1525 Budapest 114, Hungary$^{\ddag}$
\item[\mit] Massachusetts Institute of Technology, Cambridge, MA 02139, USA
\item[\panjab] Panjab University, Chandigarh 160 014, India.
\item[\debrecen] KLTE-ATOMKI, H-4010 Debrecen, Hungary$^\P$
\item[\florence] INFN Sezione di Firenze and University of Florence, 
     I-50125 Florence, Italy
\item[\cern] European Laboratory for Particle Physics, CERN, 
     CH-1211 Geneva 23, Switzerland
\item[\wl] World Laboratory, FBLJA  Project, CH-1211 Geneva 23, Switzerland
\item[\geneva] University of Geneva, CH-1211 Geneva 4, Switzerland
\item[\hefei] Chinese University of Science and Technology, USTC,
      Hefei, Anhui 230 029, China$^{\triangle}$
\item[\lausanne] University of Lausanne, CH-1015 Lausanne, Switzerland
\item[\lyon] Institut de Physique Nucl\'eaire de Lyon, 
     IN2P3-CNRS,Universit\'e Claude Bernard, 
     F-69622 Villeurbanne, France
\item[\madrid] Centro de Investigaciones Energ{\'e}ticas, 
     Medioambientales y Tecnol\'ogicas, CIEMAT, E-28040 Madrid,
     Spain${\flat}$ 
\item[\florida] Florida Institute of Technology, Melbourne, FL 32901, USA
\item[\milan] INFN-Sezione di Milano, I-20133 Milan, Italy
\item[\moscow] Institute of Theoretical and Experimental Physics, ITEP, 
     Moscow, Russia
\item[\naples] INFN-Sezione di Napoli and University of Naples, 
     I-80125 Naples, Italy
\item[\cyprus] Department of Physics, University of Cyprus,
     Nicosia, Cyprus
\item[\nymegen] University of Nijmegen and NIKHEF, 
     NL-6525 ED Nijmegen, The Netherlands
\item[\caltech] California Institute of Technology, Pasadena, CA 91125, USA
\item[\perugia] INFN-Sezione di Perugia and Universit\`a Degli 
     Studi di Perugia, I-06100 Perugia, Italy   
\item[\peters] Nuclear Physics Institute, St. Petersburg, Russia
\item[\cmu] Carnegie Mellon University, Pittsburgh, PA 15213, USA
\item[\potenza] INFN-Sezione di Napoli and University of Potenza, 
     I-85100 Potenza, Italy
\item[\prince] Princeton University, Princeton, NJ 08544, USA
\item[\riverside] University of Californa, Riverside, CA 92521, USA
\item[\rome] INFN-Sezione di Roma and University of Rome, ``La Sapienza",
     I-00185 Rome, Italy
\item[\salerno] University and INFN, Salerno, I-84100 Salerno, Italy
\item[\ucsd] University of California, San Diego, CA 92093, USA
\item[\sofia] Bulgarian Academy of Sciences, Central Lab.~of 
     Mechatronics and Instrumentation, BU-1113 Sofia, Bulgaria
\item[\korea]  The Center for High Energy Physics, 
     Kyungpook National University, 702-701 Taegu, Republic of Korea
\item[\utrecht] Utrecht University and NIKHEF, NL-3584 CB Utrecht, 
     The Netherlands
\item[\purdue] Purdue University, West Lafayette, IN 47907, USA
\item[\psinst] Paul Scherrer Institut, PSI, CH-5232 Villigen, Switzerland
\item[\zeuthen] DESY, D-15738 Zeuthen, 
     FRG
\item[\eth] Eidgen\"ossische Technische Hochschule, ETH Z\"urich,
     CH-8093 Z\"urich, Switzerland
\item[\hamburg] University of Hamburg, D-22761 Hamburg, FRG
\item[\taiwan] National Central University, Chung-Li, Taiwan, China
\item[\tsinghua] Department of Physics, National Tsing Hua University,
      Taiwan, China
\item[\S]  Supported by the German Bundesministerium 
        f\"ur Bildung, Wissenschaft, Forschung und Technologie
\item[\ddag] Supported by the Hungarian OTKA fund under contract
numbers T019181, F023259 and T024011.
\item[\P] Also supported by the Hungarian OTKA fund under contract
  number T026178.
\item[$\flat$] Supported also by the Comisi\'on Interministerial de Ciencia y 
        Tecnolog{\'\i}a.
\item[$\sharp$] Also supported by CONICET and Universidad Nacional de La Plata,
        CC 67, 1900 La Plata, Argentina.
\item[$\triangle$] Supported by the National Natural Science
  Foundation of China.
\end{list}
}
\vfill

 
\newpage
%

%


\bibliographystyle{l3stylem}

\begin{figure*}[htbp]  
\begin{center}
    \includegraphics[width=\figwidth]{fig3.epsi}
  \end{center}
  \icaption{
MSSM exclusion contours, at 95\% C.L., for the masses of a) $\serr$,
b) $\smur$, c) $\staur$ and d) $\snumt$ as a function
of the neutralino mass.
The solid and dashed lines, show the $\lambda$ and $\lambda''$
exclusion contours, respectively. The dotted line corresponds
to $\DM = 0$.
For $\DM < 0$, above this line, the exclusion contours from direct decays are shown. 
  \label{fig:fig3}}
\end{figure*}

\begin{figure*}[htbp]  
\begin{center}
    \includegraphics[width=\figwidth]{fig4.epsi}
  \end{center}
  \icaption{
MSSM exclusion contours, at 95\% C.L., for the masses of a) up-type
b) down-type scalar quarks c) $\sbottom$ and d) $\stop$ as a function
of the neutralino mass, for $\lambda''$ coupling.
The solid and dashed lines show the exclusion contours 
for a) $\qul$, $\qur$ and b) $\qdl$, $\qdr$, respectively.
For $\DM < 0$, above the dotted line, the exclusion contours from direct decays are shown. 
  \label{fig:fig4}}
\end{figure*}

\begin{figure*}[htbp]  
\begin{center}
    \includegraphics[width=\figwidth]{fig5.epsi}
  \end{center}
  \icaption{
MSSM mass limits from combined analyses.
The solid and dashed lines, labelled with the corresponding
coupling, show the
95\% C.L. lower limits on the masses of a) $\chio$, b) $\chid$ and 
c) $\slepr$, as a function of \tb, for
$0 \leq M_2 \leq 1000$ \gev{} and  $- 500$ \GeV{} $\leq \mu \leq$ 500 \GeV.
$m_0 = $ 500 \gev{} in a) and b) and $m_0 = 0$ in c). For those
values of $m_0$ the global minima on the mass limit are obtained.
  \label{fig:mlimit1}}
\end{figure*}

\end{document}